\documentclass[conference]{IEEEtran}
\IEEEoverridecommandlockouts

\usepackage{cite}
\usepackage{amsmath,amssymb,amsfonts}
\usepackage{algorithmic}
\usepackage{graphicx}
\usepackage{textcomp}
\usepackage{xcolor}
\def\BibTeX{{\rm B\kern-.05em{\sc i\kern-.025em b}\kern-.08em
    T\kern-.1667em\lower.7ex\hbox{E}\kern-.125emX}}

\usepackage{url}
\usepackage{natbib}

\title{Resolving the Correct Library: A Loader-Level Defense Solution Against Shared Object Hijacking}

\author{\IEEEauthorblockN{Can Özkan}
\IEEEauthorblockA{\textit{COSIC, KU Leuven} \\
can.ozkan@esat.kuleuven.be}
\and
\IEEEauthorblockN{Dave Singelée}
\IEEEauthorblockA{\textit{DistriNet, KU Leuven} \\
dave.singelee@kuleuven.be}
}

\date{January 2026}

\begin{document}

\maketitle

\begin{abstract}

Shared library hijacking attacks in the Linux ecosystem, including embedded Linux, are a significant concern. It fundamentally exploits the dynamic linker’s library-resolution semantics rather than modifying trusted libraries directly. Prior research has extensively analyzed attack vectors exploiting environment variables, embedded search paths, and dynamic loader internals, demonstrating that hijacking is rooted in fundamental loader behavior rather than isolated misconfigurations. Existing defenses either harden or replace the loader, enforce control-flow integrity after libraries are loaded, or apply file-centric integrity mechanisms such as signatures and measurement frameworks. However, these approaches fail to address a critical gap: none verify whether the shared object actually resolved by the loader is the intended and trusted one.

In this paper, we argue that shared library hijacking is fundamentally a loader-resolution authenticity problem and present a loader-centric verification framework that enforces authenticity guarantees for the dynamic linker’s resolution process. Our design supports both path-bound and location-independent (i.e., Build-ID–based) identity models combined with cryptographic hashing. We implement our approach on GNU libc (glibc) systems and evaluate it on both general-purpose Linux (e.g., Ubuntu) and embedded Linux (e.g., Buildroot) environments under emulation. Our results demonstrate that our proposed mechanism indeed prevents shared library hijacking attacks. 




\end{abstract}


\section{Introduction}

Modern Linux systems rely heavily on dynamically linked shared libraries to reduce memory consumption, simplify software maintenance, and enable code reuse across applications \cite{agrawal2015architectural}. Instead of statically embedding all required functionality into a binary, executables defer the resolution of external functions and symbols at runtime \cite{beazley2001inside}. During program startup, the dynamic loader (e.g., ld-linux.so in glibc-based systems) resolves the required shared objects according to a predefined search order that considers metadata embedded in ELF binaries, environment variables, system library paths, and runtime configuration files. Therefore, this mechanism provides flexibility and modularity, but also introduces a critical security boundary: the correctness of the library resolution process itself.

Shared library resolution in Linux is governed by the dynamic loader provided by the system’s C standard library. Two implementations dominate this space: glibc \cite{glibc}, used by mainstream distributions such as Ubuntu, and also deployed in many embedded Linux ecosystems, and musl \cite{musl}, which is widely adopted in embedded systems, including those based on Buildroot, Yocto, and OpenWRT. They differ in design philosophy and resolution behavior, with glibc focusing on flexibility and musl favoring simplicity. These differences directly influence the library resolution order and attack surface, showing that shared library hijacking is a loader-centric issue rather than a distribution-specific one.


In Linux, shared objects are typically identified by their SONAME (Shared Object Name), such as libssl.so or libcrypto.so. When an executable starts, the dynamic loader searches for libraries matching these names across multiple candidate directories. The exact search order depends on the loader implementation (e.g., glibc or musl). In glibc-based systems, resolution may first consider \texttt{DT\_RPATH} entries (when \texttt{DT\_RUNPATH} is absent), followed by paths specified through the \texttt{LD\_LIBRARY\_PATH} environment variable, \texttt{DT\_RUNPATH} entries, the dynamic linker cache (\texttt{/etc/ld.so.cache}), and finally default system directories such as \texttt{/lib} and \texttt{/usr/lib}~\cite{resolutionorder}. The loader ultimately selects the first matching library encountered during this resolution process and maps it into the process address space.

Fundamentally, shared-object hijacking attacks exploit this resolution behavior by manipulating library search precedence. An attacker introduces a malicious shared object that uses the same filename or SONAME as a legitimate dependency, instead of modifying trusted binaries or tampering with legitimate system libraries. In other words, the attacker causes the dynamic loader to transparently resolve and load the attacker-controlled object by ensuring that the malicious library resides in a directory searched before the legitimate location. Common attack vectors include manipulating LD\_LIBRARY\_PATH, abusing writable directories referenced in RPATH or RUNPATH, placing malicious libraries in application-local directories, or exploiting unsafe runtime loading patterns involving dlopen().

For example, suppose that an application depends on libfoo.so. Under normal conditions, the loader resolves the legitimate library from /usr/lib/libfoo.so. However, if an attacker can place a malicious libfoo.so in a user-controlled directory referenced by, for example, LD\_LIBRARY\_PATH, the loader may resolve the attacker-controlled library first. Since the malicious library exports the expected symbols and interfaces, the application continues to operate normally while executing attacker-supplied code inside the victim process context. More importantly, this attack does not require memory corruption, binary patching, or modification of the original library. The attack succeeds solely because the loader resolves the wrong object.

This property makes shared library hijacking fundamentally different from traditional integrity violations. Traditional integrity mechanisms typically verify whether a file has been modified or corrupted. In contrast, hijacking attacks often involve entirely valid and well-formed ELF shared objects that have never been tampered with after creation. The problem is therefore not file integrity, but resolution correctness: the loader successfully loads a legitimate ELF object, but not the one originally intended by the application or system designer.

The consequences of successful hijacking attacks are severe because malicious code executes before most application-level defenses become active. Once loaded, the attacker-controlled library executes with the privileges and security context of the target process. This enables arbitrary code execution, stealthy persistence, credential theft, bypassing of application logic, and local privilege escalation when privileged binaries are targeted. In embedded and industrial Linux environments, the impact can be even more significant, potentially enabling persistent compromise of firmware components, manipulation of control logic, or interference with safety-critical operations.

Despite extensive prior work on shared library hijacking, existing defenses fail to address its root cause. Loader hardening techniques protect the internal linker state or restrict the influence of environment variables, but still allow incorrect libraries to be resolved. Control-flow and GOT/PLT protections assume that the intended library has already been loaded, whereas integrity mechanisms, such as signature verification and IMA\cite{17_IMA}, only ensure that a loaded library is unmodified, not that it is the correct one. Compartmentalization and sandboxing approaches reduce post-compromise impact but accept hijacking as inevitable.

As our contribution, we introduce a loader-centric glibc prevention mechanism that enforces checking shared library identity at load time by binding dependencies to immutable build identifiers and cryptographic hashes. Specifically, our solution controls which libraries are loaded at the glibc dynamic loader level (i.e., by checking each library to be loaded against a white list of hash values at the loader). This prevents hijacking attacks that rely on precedence manipulation.

\section{Related Work}
Shared library hijacking has been quite extensively studied in the Linux ecosystem, with research spanning from early secure programming guidelines to recent analyses of dynamic linker architecture. 


\subsection{Attacks on Dynamic Linking}
Early work established that the flexibility of ELF dynamic linking introduces exploitable resolution semantics. Wheeler \cite{5_wheeler2001secure} documented how attackers can abuse LD\_PRELOAD and LD\_LIBRARY\_PATH to interpose malicious libraries. This lays the foundation for later exploitation techniques. These vectors were later demonstrated in real-world malware and rootkits, such as Jynx2, which leveraged LD\_PRELOAD and /etc/ld.so.preload for persistent userland compromise \cite{8_carbone2014malware}.

Subsequent research expanded the attack surface by analyzing RPATH and RUNPATH misuse, particularly when combined with the \$ORIGIN variable. This shows that embedded search paths can be exploited even when environment variables are restricted \cite{1_payer2012safe, 4_williams2021improving, 10_zakaria2025exploiting}. These weaknesses, therefore, resulted in multiple CVEs and demonstrated that misconfigured binaries remain vulnerable regardless of runtime protections.

More advanced work has shown that modern mitigations are insufficient when attackers exploit the dynamic loader's internals directly. Federico et al. introduced the \textit{Leakless} technique, demonstrating that RELRO can be bypassed by manipulating ELF metadata and dynamic linker state \cite{6_di2015elf}. Ge et al. identified copy relocation violations (\textit{COREV}), showing that the loader may relocate read-only data into writable memory. This undermines CFI assumptions and enables control-flow hijacking \cite{2_ge2017evil}. Additional techniques bypass context-based checks by forging loader execution contexts \cite{13_ditullio2020context}. These works show that shared library hijacking is not limited to misconfiguration but is rooted in fundamental loader behavior.

\subsection{Loader- and Runtime-Level Defenses}
To counter these attacks, several defenses modify or harden the dynamic loader. TRuE replaces the standard loader with a security-aware runtime that restricts the influence of environment variables, isolates loader state, and reduces indirect control-flow transfers \cite{1_payer2012safe}. While effective against many attacks, such approaches require significant system changes and face deployment barriers.

Other defenses focus on protecting Global Offset Table (GOT) and Procedural Linkage Table (PLT) structures. SCC dynamically relocates the GOT and validates PLT calls to prevent unauthorized redirection \cite{7_park2005new}, while SecGOT and related CFI-based approaches enforce control-flow constraints on indirect calls \cite{22_zhang2013secgot, 24_jeong2020cfi, 26_zhang2015control}. These techniques are effective against memory corruption attacks targeting loaded libraries, but assume that the correct library has already been resolved.

\subsection{Integrity Enforcement and Monitoring}

Integrity-based mechanisms aim to ensure that loaded libraries have not been tampered with. The Linux Integrity Measurement Architecture (IMA)-like approaches and signature-based verification (such as Extended Verification Module (EVM)) prevent modified or unsigned libraries from being loaded \cite{17_IMA}. However, integrity enforcement is file-centric. More specifically, it verifies authenticity, not resolution correctness (which we will explain in detail later). An attacker can substitute a different, but unmodified, library with the same name, thereby bypassing integrity checks entirely.

Runtime monitoring and detection systems attempt to identify unsafe loading behavior. Kwon et al. proposed dynamic instrumentation to detect unsafe component loadings, demonstrating that such vulnerabilities are widespread in practice \cite{14_kwon2011automatic}. Additional runtime tools monitor environment variables, RUNPATH usage, and anomalous loading behavior \cite{9_vijayakumar2014protecting, 27_fu2012dynamic}. While valuable for detection and forensics, these approaches are reactive and cannot prevent exploitation at \textit{load time}.

\subsection{Compartmentalization and Architectural Approaches}
More recent work explores limiting the impact of hijacked or vulnerable libraries by isolating them. LibVM sandboxes libraries in isolated execution environments \cite{23_goonasekera2015libvm}, while dynamic library compartmentalization further separates libraries into security domains \cite{16_larose2023dynamic}. Related approaches reduce attack surface through debloating \cite{30_porter2020blankit} or thread-level isolation \cite{29_rommel2023thread}. These defenses assume compromise and focus on containment rather than prevention.

Other work argues that the dynamic linker itself is a fundamental architectural bottleneck. Castes et al. describe the linker as the “narrow waist” of operating systems and say that its current design inherently trades security for flexibility \cite{15_castes2023dynamic}. Hardware-assisted isolation \cite{19_silakov2012using} and information-flow protections \cite{18_lu2017securing} have also been explored, but they require specialized support and do not directly address resolution semantics.

\subsection{Provenance, Analysis, and Dual-Purpose Studies}
Several studies bridge the gap between attack and defense by analyzing loader behavior while proposing mitigations. Lprov tracks library provenance to enable forensic analysis of anomalous loading behavior \cite{28_wang2018lprov}. Other work combines real-time loading control with code reuse defenses \cite{25_jianjun2022defense}. 

Despite the breadth of existing work, a critical gap remains. None of the surveyed defenses enforces resolution correctness, namely, the guarantee that the dynamic linker loads the specific library instance intended by the binary. Loader hardening protects internal state \cite{1_payer2012safe}, GOT/PLT defenses protect control flow \cite{7_park2005new, 22_zhang2013secgot}, integrity mechanisms protect file contents \cite{17_IMA}, and compartmentalization limits damage \cite{16_larose2023dynamic, 23_goonasekera2015libvm}. Yet, none bind a binary’s dependencies to a verifiable library identity.
Furthermore, while unsafe loading patterns are well documented \cite{14_kwon2011automatic}, there is a lack of practical static analysis tools that enable analysts to identify potential entry points in binaries without proactively executing them. Existing approaches are either dynamic, research prototypes, or require manual operations.

\subsection{Research Gap}
Our work, therefore, directly addresses these issues. We constrain the dynamic linker’s resolution semantics, the root cause exploited by hijacking attacks, by enforcing shared library identity at load time using immutable build identifiers and cryptographic hashes. This approach complements existing integrity and loader defenses and seeks to answer the key question that current state-of-the-art solutions do not: Is a shared object library being loaded the correct library?


\section{Background}
\subsection{glibc vs. musl}


In practice, both general-purpose and embedded Linux ecosystems converge on a small set of C library implementations. Ubuntu and most desktop and server distributions use glibc by default, making it the de facto standard in enterprise and research environments. Embedded build systems such as Buildroot and Yocto, however, support multiple libc implementations, most notably glibc and musl, while OpenWRT traditionally relies on musl and supports glibc. Historically, uClibc and uClibc-ng have also been used in embedded systems to minimize footprint, though their adoption has decreased in favor of musl due to maintenance and security considerations. As a result, glibc and musl cover the vast majority of contemporary Linux systems across both general-purpose and embedded domains.

glibc emphasizes compatibility, extensibility, and feature completeness. Its dynamic loader implements a rich and highly flexible resolution algorithm, supporting RPATH, RUNPATH, \$ORIGIN, environment-variable–controlled search paths, dynamic auditing (LD\_AUDIT), and interposition mechanisms such as LD\_PRELOAD. 

Musl, by contrast, adopts a minimalist design philosophy focused on simplicity, resource efficiency, attention to correctness, safety under resource exhaustion, and ease of deployment \cite{musl}. Its loader omits several extensibility features present in glibc and enforces a more deterministic resolution process. However, musl still supports RPATH, RUNPATH, and \$ORIGIN, and therefore remains susceptible to hijacking attacks that manipulate resolution precedence via embedded search paths. uClibc exhibits similar limitations: while smaller and simpler than glibc, it does not fundamentally enforce resolution correctness and inherits the same class of vulnerabilities at the semantic level.


\subsection{BuildID}

Modern Linux toolchains commonly embed a Build-ID into ELF executables and shared objects to uniquely identify a specific build artifact. The Build-ID is stored as an ELF note within the \texttt{PT\_NOTE} program header, typically using the \texttt{NT\_GNU\_BUILD\_ID} note type defined by the GNU toolchain ecosystem~\cite{elfspec,gnuldbuildid}. During linking, the linker generates this identifier and embeds it into the final ELF object.

The Build-ID primarily serves as a stable artifact identifier, not a cryptographic integrity mechanism. In practice, Build-IDs are widely used by debugging infrastructures, symbol servers, crash analysis frameworks, and package management systems to associate binaries with debugging symbols and metadata. Since the identifier is embedded in the ELF object itself, it remains independent of the binary's file name or filesystem location. Therefore, renaming or relocating a shared object does not affect its Build-ID value.
\section{Adversary Model and Trust Assumptions}
\label{sec:threat_model}

Shared-object hijacking attacks exploit the dynamic loader’s resolution order. An attacker places a malicious library with the same name as a legitimate dependency in a directory that is searched earlier by the loader (e.g., via manipulated environment variables such as \texttt{LD\_LIBRARY\_PATH}, writable directories in the search path, or misconfigured application-local paths), and the loader resolves it before the legitimate one. 


Knowing this, it is essential to secure the application’s dependencies and verify the correctness of the shared objects it loads. Our objective is to ensure that all applications and system modules are explicitly approved. Furthermore, instead of verifying whether a file has been modified (which is ineffective at mitigating hijacking attacks because the adversary does not tamper with the original library), we verify whether the resolved shared object is the one the system intends to allow for that application.


Our threat model focuses on shared-object hijacking attacks that exploit dynamic loader resolution semantics at runtime. The adversary’s objective is to cause a target application or system component to load an attacker-controlled shared object in place of the legitimate dependency intended by the system.

We assume an adversary capable of influencing the dynamic loader’s library-resolution process via standard user-space mechanisms. Specifically, the adversary may:

\begin{itemize}
    \item place malicious shared objects in the user-space writable filesystem,
    \item manipulate library search precedence via environment variables such as \texttt{LD\_LIBRARY\_PATH},
    \item exploit unsafe \texttt{RPATH} or \texttt{RUNPATH} configurations,
    \item write to application library directories, and
    \item trigger runtime loading through mechanisms such as \texttt{dlopen()}.
\end{itemize}

The attacker may therefore introduce alternative ELF shared objects that export the expected symbols and interfaces of legitimate dependencies. However, the adversary does not modify the original trusted library itself. Instead, the attack succeeds by causing the dynamic loader to resolve a different object during runtime dependency resolution.

We assume that the attacker lacks administrative or kernel-level privileges (i.e., (s)he only has user-space privileges) and cannot compromise the trusted computing base of the verification system. In particular, the adversary cannot:

\begin{itemize}
    \item modify the verifier implementation,
    \item alter the authenticated allowlist manifest,
    \item replace the cryptographic public key used for signature verification,
    \item tamper with the trusted provisioning environment,
    \item or bypass the cryptographic primitives used for integrity validation.
\end{itemize}

We further assume that the operating system kernel, the underlying filesystem integrity of trusted system directories, and the cryptographic verification routines remain trustworthy throughout execution.

Our mechanism is designed to prevent unauthorized runtime dependency substitution after deployment. Consequently, attacks that compromise the software supply chain prior to provisioning are outside the scope of this work. For example, if an attacker compromises the build pipeline and produces malicious shared objects that are subsequently signed and approved during provisioning, the loader-level verification mechanism will accept those artifacts as valid. Similarly, attacks involving full-system compromise, kernel compromise, or the privileged modification of the verifier itself are considered out of scope.

\section{Loader-Centric Enforcement for Shared Objects}

\subsection{Application vs. System Modules}

Before we discuss our proposed solution, it is important to distinguish between \emph{application modules} and \emph{system modules}.


Application-level libraries are shared objects explicitly declared as dependencies of an executable or another shared object. They are recorded in the ELF dynamic section (DT\_NEEDED entries) \cite{elfspec} and are resolved by the dynamic loader during process initialization. For example, when executing /usr/bin/curl, libraries such as \textit{libcurl.so}, \textit{libssl.so}, and \textit{libc.so.6} appear as declared libraries that are loaded at runtime, i.e., dependencies of the application.


System-level runtime libraries are shared objects that are not declared in the application's DT\_NEEDED entries but are loaded dynamically at runtime by system components such as glibc. A prominent example is the Name Service Switch (NSS) mechanism. When performing hostname resolution, glibc consults /etc/nsswitch.conf and may dynamically load modules such as: \textit{/usr/lib/x86\_64-linux-gnu/libnss\_mdns4\_minimal.so.2} These modules are not linked into the application binary and are therefore not visible in static dependency analysis (e.g., ldd). Instead, they are loaded via internal dlopen() calls within libc, depending on the system configuration. Therefore, we also propose a helper script that edits the manifest file.

Our proposed enforcement mechanism, discussed later in the paper, operates within the dynamic loader itself. As a result, any shared object resolved by the linker, regardless of origin, is subject to verification. This includes: i) libraries declared via DT\_NEEDED, ii) libraries introduced via dlopen(), iii) system-level modules loaded implicitly by libc (e.g., NSS modules). Therefore, to account for runtime-discovered dependencies (e.g., NSS modules and \texttt{dlopen()}-loaded libraries) that are not visible during static dependency analysis, we additionally provide a helper script that updates the
authenticated manifest with newly observed legitimate shared objects.

Since our enforcement operates at the dynamic loader boundary, it verifies all shared objects regardless of how they are introduced. This ensures comprehensive, full-system shared library protection, not only application-scoped integrity enforcement.

\subsection{Main Concept of Our Design}

Fundamentally, our proposed solution is an allow list (that we will refer to as a manifest) that enumerates which shared objects should be loaded. Each entry in the allowlist consists of a primary key and a digest; we call this the manifest of that file. A central design choice is how to identify shared objects in a manner that is robust to relocation and renaming, yet simple enough for practical deployment. For this, we considered three alternatives.

The simplest option (i.e., option A) is to key integrity decisions by the resolved absolute path of the shared object. Although straightforward, this approach is fragile: any change to the file system layout, relocation of libraries, or use of containerized or read-only root file systems requires regenerating the allow list. That is, a layout change requires a new manifest file.

A more flexible approach (i.e., option B) is to key by SONAME and hash the file contents. This avoids absolute path dependence and aligns more closely with dynamic linking semantics. However, multiple distinct builds can legitimately share the same SONAME, requiring additional disambiguation and policy complexity. For instance, lib1.so.1.2 and lib1.so.1.3 shared the same SONAME - lib1.so.1.

The third approach, which keys allowlist entries by ELF Build-ID, combines with a cryptographic hash of the file contents. The Build-ID, stored in the \texttt{NT\_GNU\_BUILD\_ID} ELF note, uniquely identifies a specific build artifact independent of its filename or location \cite{elfspec}. In our design, the Build-ID serves solely as an identity key, while integrity is enforced by comparing the SHA-256 hash of the loaded file against the expected hash bound to that Build-ID. This clean separation between identity and integrity avoids reliance on path-based heuristics and ambiguous naming. For instance, lib1.so.1.2 and lib1.so.1.3 have different Build-IDs, even though they share the same SONAME. Therefore, we use Build-IDs to distinguish libraries.

We therefore implemented and evaluated two loader-level enforcement variants. Both variants enforce cryptographic integrity via SHA-256 hashing, but they differ in how the library identity is established:

\textbf{Variant A (Path-Based Identity):} Shared objects are identified by their canonical filesystem path combined with a SHA-256 digest. This approach enforces that a specific file at a specific location must match an approved cryptographic hash. This approach is useful when no Build-ID value is attached to the binary to be protected.

\textbf{Variant B (Build-ID-Based Identity):} Shared objects are identified by their ELF Build-ID combined with a SHA-256 digest. The Build-ID serves as a location-independent identity key, while SHA-256 enforces full byte-level integrity.

The two variants represent different trust-binding strategies. Path-based identity provides strict location binding and is conceptually simpler. Build-ID-based identity provides relocation resilience and decouples trust from filesystem layout. In both cases, SHA-256 is required for integrity enforcement.

\subsection{Implementation on glibc}

We implement both variants on glibc-based systems using the \texttt{LD\_AUDIT} interface \cite{rtldaudit} provided by the dynamic loader. This interface enables an auditing module to observe every shared object that the loader maps, including transitive dependencies (i.e., dependencies that depend on other dependencies), objects loaded explicitly via \texttt{dlopen()}\ and system modules.

At process startup, the auditing module verifies a cryptographic signature over the allow list manifest using an Ed25519 public key provisioned as part of the trusted system image. Depending on the selected enforcement mode, the manifest maps either: i) canonical library paths to expected SHA-256 hashes (Path Mode), or ii) build-IDs to expected SHA-256 hashes (Build-ID Mode). The verifier then extracts either the canonical path or the Build-ID at load time and performs integrity validation accordingly. Once authenticated, the manifest is loaded into memory. For each shared object load event, the module extracts the Build-ID from the ELF note section, computes the SHA-256 hash of the resolved file, and compares the result to the expected value. Any mismatch or missing entry results in immediate termination of the process before execution returns to the application.


\section{Our Solution}
\label{subsec:manifest-sign-enforce}

Our approach consists of two phases: an offline provisioning phase that generates authenticated reference values for approved shared objects, and an online enforcement phase that verifies every relevant DSO at load time. We also assume that the attacker is only present during the online phase. This subsection describes each step in detail, including generating the allowlist (the manifest), signing the manifest, and enforcing the policy by the verifier. Below, we explain the process for Build-ID-based enforcement. Since both enforcement variants follow the same provisioning, signing, and verification workflow, we describe only the Build-ID-based variant in detail. The only difference between the two approaches is the manifest lookup key: the path-based variant verifies using canonical filesystem paths, whereas the Build-ID-based variant uses ELF Build-IDs.


\subsection{Offline Provisioning: Manifest Generation and Signing}
\label{subsec:offline-provisioning}

Provisioning is performed in a trusted build environment (e.g., the developer machine, CI pipeline, or firmware build host). The input is a set of approved shared objects (DSOs) that are intended to be loadable at runtime. In our prototype, these are libraries and plugins shipped with the program. The output is an allowlist manifest and an accompanying signature. The private signing key never leaves the provisioning environment; only the public verification key is deployed to the target system.

\paragraph{Step 1: Ensuring Build-ID availability}
In our proposal, we use ELF Build-ID as a primary key in the allowlist. Therefore, each DSO must contain a \texttt{NT\_GNU\_BUILD\_ID} note. In modern toolchains, this is typically enabled by default, but we explicitly compile our DSOs with \texttt{-Wl,--build-id} to ensure the note is present. During provisioning, we verify that a Build-ID exists; if not, the artifact is rejected from the allowlist generation process. This check prevents ambiguous policy entries and ensures stable artifact identification across relocations and renames.

\paragraph{Step 2: Extracting the Build-ID (artifact identity)}
For each DSO, we extract the Build-ID from its ELF notes. This Build-ID functions purely as an identifier: it selects which allowlist entry maps to the loaded object. It avoids reliance on absolute paths or SONAME-only naming. Furthermore, while the manifest generation process computes the transitive closure of dependencies using \texttt{ldd}, this approach captures only libraries declared through \texttt{DT\_NEEDED} entries. Some shared objects may be loaded dynamically at runtime (e.g., via \texttt{dlopen()}) and therefore do not appear in the static dependency graph. In such cases, additional legitimate libraries can be incorporated into the allowlist by re-running our helper script against the target application. The script updates the manifest by enumerating the observed libraries, appending their canonical paths, Build-IDs, and SHA-256 hashes, and then re-signing it. This update mechanism enables controlled extension of the allowlist and preserves the integrity and authenticity guarantees enforced by the verifier.

\paragraph{Step 3: Computing the SHA-256 hash (integrity measurement)}
For each DSO, we compute the SHA-256 hash of the entire on-disk byte sequence of the shared object file. Hashing is performed in a streaming manner: the file is read in fixed-size chunks (e.g., 1~MB), and each chunk is fed into the SHA-256 state until the end of the file is reached. The resulting 32-byte digest is hex-encoded. This procedure provides a byte-for-byte integrity check of the artifact; any modification, corruption, or substitution, therefore, results in a mismatching digest at runtime.

\paragraph{Step 4: Building the hash list and manifest file}
The hash list is constructed as a set of pairs \(\langle \texttt{Build-ID}, \texttt{SHA256} \rangle\) for all approved DSOs. We serialize this set into a plain-text manifest file, one entry per line, using the format:
\[
\texttt{<library\_path><build\_id\_hex> <sha256\_hex>}
\]
The simplicity of this format supports easy inspection and reproducibility. It also enables deterministic lookup: at runtime, the verifier extracts the Build-ID from the loaded object and performs a direct match against manifest entries.

\paragraph{Step 5: Key pair generation (Ed25519).}
To authenticate the manifest, we generate an Ed25519 signing key pair during provisioning. The private key is stored only in the provisioning environment (e.g., \texttt{ed25519\_priv.pem}). The corresponding public key (\texttt{pub.pem}) is exported and deployed to the target system. This separation ensures that an attacker who compromises the target filesystem cannot forge a valid manifest update.

\paragraph{Step 6: Signing the manifest.}
We compute a digital signature over the \emph{exact bytes} of the manifest file. More specifically, the signing operation takes \texttt{manifest.txt} as input and produces a detached signature file \texttt{manifest.sig}. In our prototype, we use OpenSSL’s Ed25519 support with a raw signing operation so that signature verification corresponds to a one-shot Ed25519 verification over the manifest’s byte stream. Since the signature covers the exact manifest bytes, any modification to the manifest entries (e.g., addition, deletion, reordering, or whitespace changes) invalidates the signature. This property is critical. It prevents attackers from editing allowlist entries even if they can write to the manifest location.

\paragraph{Provisioning output.}
The provisioning phase produces three artifacts: (i) the manifest file \texttt{manifest.txt} containing the allowlist entries, (ii) the attached signature \texttt{manifest.sig}, and (iii) the Ed25519 public key \texttt{pub.pem}. These artifacts are deployed to a trusted directory on the target (e.g., \texttt{/etc/dsoverify/}). We further assume that these files are not writable by attackers not having a root access; a common misconfiguration is to make them world-writable.

\subsection{Online Enforcement: Load-Time Verification}
\label{subsec:online-enforcement}

Upon successful provisioning, we then enforce the policy using the dynamic loader’s auditing interface \texttt{LD\_AUDIT} on glibc. When the application starts with the audit module enabled, the verifier executes an initialization routine before the program’s main logic runs. During initialization, the verifier reads \texttt{manifest.txt}, \texttt{manifest.sig}, and \texttt{pub.pem} from the trusted directory and verifies the signature. If signature verification fails, enforcement terminates execution immediately. If it succeeds, the verifier parses \texttt{manifest.txt} into an in-memory mapping from Build-ID to expected SHA-256 hash values.

Enforcement occurs in the loader callback, which is triggered whenever a shared object is loaded. For each load event, the verifier obtains the resolved filesystem path of the DSO from the loader. This path is used only to locate the file for parsing and hashing; it is not used as a policy key. Below is what happens every time a library is loaded.

\paragraph{Step 1: Extract Build-ID of the loaded object.}
The verifier parses the ELF object and extracts the \texttt{NT\_GNU\_BUILD\_ID} note from the \texttt{PT\_NOTE} program headers \cite{elfspec}. If the object does not contain a Build-ID, the verifier denies execution, and the program halts. This choice prevents an attacker from bypassing identification by stripping Build-ID or using non-standard artifacts. 


\paragraph{Step 2: Compute SHA-256 of the loaded object.}
The verifier computes the SHA-256 of the loaded object file on disk using the same streaming procedure as in provisioning. This produces a digest that should match the expected reference value if and only if the artifact is exactly the approved version.

\paragraph{Step 3: Lookup and comparison.}
Using the extracted Build-ID, the verifier performs a lookup in the in-memory manifest map. If the Build-ID is not present, the object is not approved and the verifier denies execution. If the Build-ID is present, the verifier compares the computed digest against the manifest’s expected digest. A mismatch indicates that either the artifact has been modified or that an attacker has substituted a different object while preserving naming. In both cases, execution is denied.

\begin{figure*}[t]
    \centering
    \includegraphics[width=0.8\linewidth]{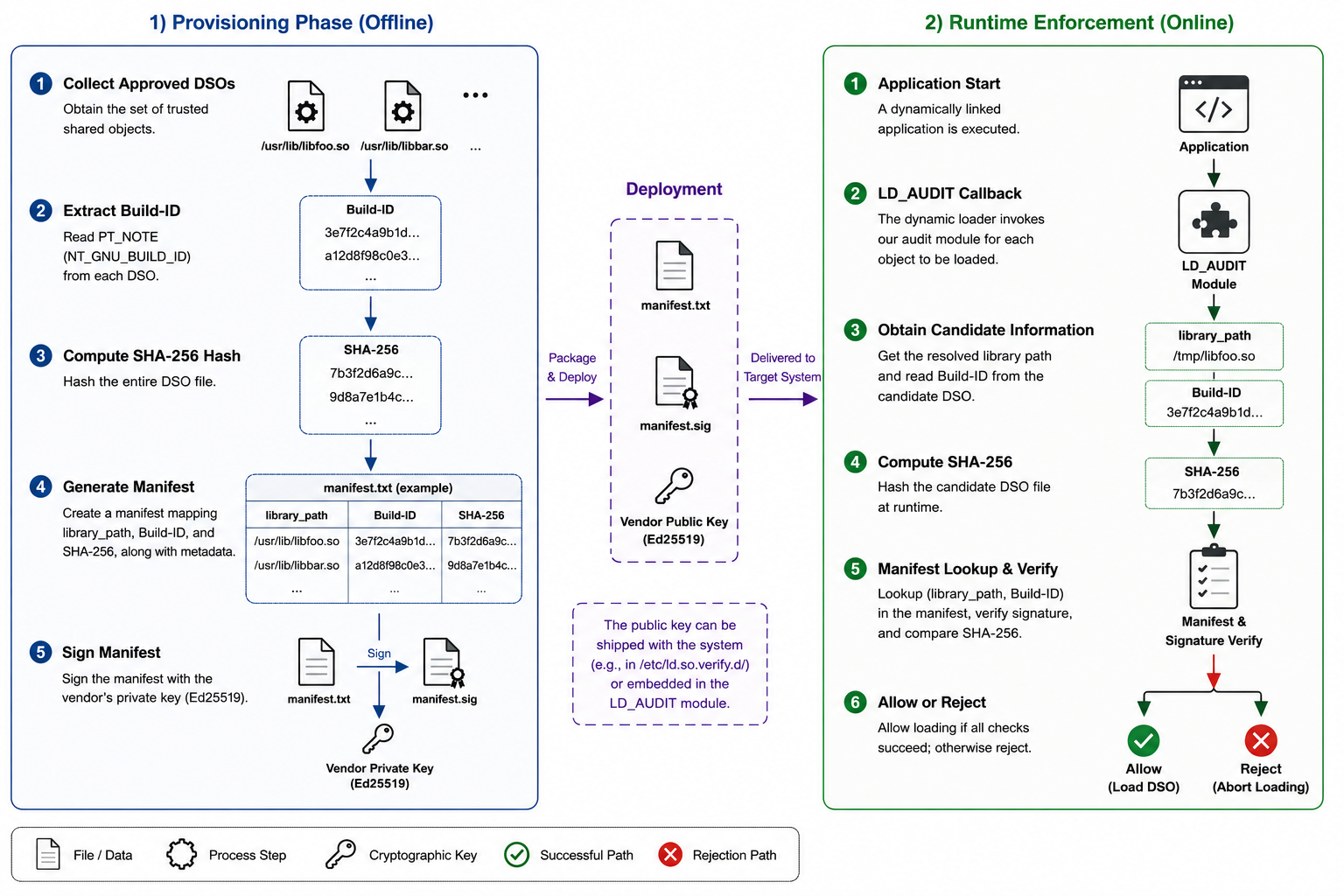}
    \caption{Proposed offline provisioning and online runtime enforcement workflow}
    \label{fig:workflow}
\end{figure*}

Figure~\ref{fig:workflow} presents the overall provisioning and runtime enforcement workflow of the proposed verification framework, including manifest generation, signature creation, and load-time validation.


\subsection{Intuition behind our solution}
The enforcement strategy we employ specifically addresses the shared object hijacking threat model in which attackers place malicious objects earlier in the search order rather than overwriting legitimate libraries. Under such attacks, the loader resolves a different DSO than intended, but the verifier treats this as an unapproved artifact because its Build-ID is not allowlisted or its hash does not match. Therefore, even though the malicious library may be a valid ELF object and unmodified, it is rejected because it is not part of the application's approved dependency set.

\section{Performance Evaluation}

This section evaluates the runtime overhead introduced by the proposed shared library verification mechanism. We benchmark two widely deployed software applications: \texttt{curl} and \texttt{OpenSSL}. The evaluation compares three configurations:

\begin{itemize}
    \item Baseline (no verification)
    \item Path-based verification (canonical path + SHA-256)
    \item Build-ID-based verification (Build-ID + SHA-256)
\end{itemize}

All experiments were performed on Ubuntu 24.04 using \texttt{hyperfine} \cite{hyperfine} with 20 warm-up runs and 200 measured executions per configuration. Each reported value corresponds to the mean execution time with standard deviation.

\subsection{curl}

The \texttt{curl} binary declares 33 shared object dependencies as reported by \texttt{ldd}. Each library is verified at load time under strict full-system enforcement.

The workload used for benchmarking was:

\begin{verbatim}
curl -sS -o /dev/null file:///etc/hosts
\end{verbatim}

This command ensures deterministic execution without network variability.

\begin{table}[h]
\centering
\caption{Performance results for \texttt{curl} (33 DSOs)}
\begin{tabular}{lccc}
\hline
Configuration & Mean (ms) & Std Dev (ms) & Overhead (ms) \\
\hline
Baseline & 9.8 & 2.1 & -- \\
Path-based & 112.6 & 10.5 & 102.8 \\
Build-ID-based & 110.9 & 6.6 & 101.1 \\
\hline
\end{tabular}
\end{table}

The absolute overhead is approximately 101--103 ms. Dividing by the 33 verified libraries yields an average verification cost of approximately 3.06 ms per library. The negligible difference between the Path-based and Build-ID-based configurations indicates that SHA-256 hashing dominates the verification cost, while Build-ID extraction introduces minimal additional overhead.

\subsection{OpenSSL}

The \texttt{openssl} binary declares 5 shared object dependencies according to \texttt{ldd}. The benchmarked workload was:

\begin{verbatim}
openssl list -digest-algorithms > /dev/null
\end{verbatim}

This workload triggers initialization of cryptographic subsystems while remaining deterministic.

\begin{table}[h]
\centering
\caption{Performance results for \texttt{OpenSSL} (5 DSOs)}
\begin{tabular}{lccc}
\hline
Configuration & Mean (ms) & Std Dev (ms) & Overhead (ms) \\
\hline
Baseline & 5.4 & 1.5 & -- \\
Path-based & 49.9 & 4.0 & 44.5 \\
Build-ID-based & 49.3 & 3.3 & 43.9 \\
\hline
\end{tabular}
\end{table}

The observed overhead is approximately 44 ms. When normalized over 5 verified libraries, this corresponds to roughly 8.8 ms per library. Note that this value is higher than the per-library normalized cost in the case of curl. This can be explained by a fixed initialization overhead, which has a smaller impact when distributed over more dependencies. 

Across both applications, the evaluation confirms:

\begin{itemize}
    \item The results are consistent with linear scaling with respect to the number of verified DSOs.
    \item SHA-256 hashing is the dominant cost component.
    \item Build-ID-based enforcement introduces negligible additional overhead compared to path-based enforcement.
\end{itemize}


\subsection{Interpretation of Startup Overhead}

The measured relative slowdowns appear significant because the evaluated workloads consist of short-lived utilities with extremely small baseline execution times. Applications such as curl and openssl in our benchmarks typically complete execution within only a few milliseconds under normal conditions. Therefore, the fixed verification cost introduced during process initialization dominates total runtime and intensifies the observed relative overhead.

However, the proposed mechanism performs verification only during shared object loading at process startup. Once libraries have been authenticated and mapped into memory, no additional runtime instrumentation or continuous monitoring is performed during steady-state execution. As a result, the overhead is incurred only once per process lifetime.

This distinction is particularly important for long-running applications and embedded services. Daemons, servers, industrial control applications, and continuously running user-space services typically remain active for extended periods after initialization. In such deployment scenarios, the startup verification cost is amortized over the process's operational lifetime, substantially reducing its practical impact. For these classes of applications, the measured overhead therefore represents a one-time initialization latency, not a sustained runtime performance penalty.

The results also show that verification overhead scales approximately linearly with the number of dynamically loaded shared objects. This behavior is expected because each library requires cryptographic hashing and identity verification during loading. While dependency-heavy applications may therefore experience higher startup latency, the mechanism introduces no measurable overhead after initialization completes. Therefore, we suggest that our proposed approach is most appropriate for environments where startup latency is less critical than runtime integrity guarantees, such as embedded Linux systems, long-running services, appliance-style deployments, and security-sensitive workloads.

\section{Security Evaluation}

We analyze the security of the proposed mechanism with respect to the adversary model defined in Section~\ref{sec:threat_model}. The attacker can influence the dynamic loader's resolution process by placing malicious shared objects in writable user-space locations, manipulating search precedence through environment variables, abusing unsafe \texttt{RPATH} or \texttt{RUNPATH} configurations, and triggering runtime loading through \texttt{dlopen()}. However, the attacker cannot modify the verifier, alter the signed manifest, replace the public verification key, compromise the trusted provisioning environment, or bypass the cryptographic primitives used for signature and hash verification.

Under these assumptions, a shared object can be loaded only if it satisfies the verifier's policy. During provisioning, each approved DSO is represented in the manifest by its resolved library path, ELF Build-ID, and SHA-256 digest. The manifest is signed using an Ed25519 private key that remains outside the target system. At runtime, the verifier first authenticates the manifest. If signature verification fails, execution is terminated. Therefore, an attacker who can write to the filesystem but cannot forge the manifest signature cannot add a malicious library to the allowlist, remove protected entries, or modify expected hash values.

We next consider search-order hijacking attacks. In these attacks, the adversary places a malicious shared object earlier in the loader's search order so that it is resolved before the legitimate dependency. Although the dynamic loader may select the attacker-controlled object, the verifier observes the resolved object before application-level execution proceeds. If path-based enforcement is used, the malicious object is rejected because its canonical path does not match the approved path in the manifest. If Build-ID-based enforcement is used, the object is rejected unless its Build-ID appears in the authenticated manifest and its SHA-256 digest matches the approved digest. Therefore, simply reusing the same filename or SONAME as a legitimate dependency is insufficient to bypass enforcement.

We also consider malicious objects that attempt to imitate legitimate dependencies by exporting the expected symbols and preserving ABI compatibility. Such objects may allow the target application to start normally in an unprotected system. However, symbol compatibility is irrelevant to the verifier's decision. The verifier authenticates the resolved DSO using manifest-bound identity and byte-level integrity checks before accepting the object as an approved dependency. Therefore, an attacker-controlled library with compatible exported symbols is still rejected unless it is explicitly approved during provisioning.

Our mechanism also covers dynamically loaded libraries. Since enforcement is placed at the dynamic loader boundary, it applies not only to dependencies declared through \texttt{DT\_NEEDED}, but also to objects that are introduced through \texttt{dlopen()} and system-level modules loaded implicitly by libc. Thus, an attacker cannot evade verification just by delaying loading until runtime or by targeting transitive or system-level shared objects.

The Build-ID is used only as an identity key and not as a standalone integrity guarantee. If an attacker modifies an approved shared object and preserves its original Build-ID, the SHA-256 digest computed at load time will differ from the digest stored in the authenticated manifest, and execution will be denied. Conversely, if an attacker creates a different shared object with a new Build-ID, the Build-ID lookup will fail unless that object was approved during provisioning. Thus, the combination of Build-ID and SHA-256 protects against both dependency substitution and post-build binary modification.

The proposed mechanism does not protect against attacks outside the stated trust assumptions. If an attacker obtains administrative privileges, and therefore modifies the verifier, replaces the public key, tampers with the trusted provisioning process, or compromises the build pipeline before manifest generation, then malicious artifacts may be approved and accepted at runtime. Similarly, attacks against the operating system kernel, the cryptographic implementation, or the loader mechanism itself are outside the scope of this work. Within the stated adversary model, however, any attacker-controlled shared object introduced through search-order manipulation will fail at least one of the required checks: manifest authentication, identity lookup, or SHA-256 digest verification.


Additionally, to empirically validate our proposed mechanism, we conducted shared-library-hijacking experiments that emulate common search-order manipulation attacks using \texttt{LD\_LIBRARY\_PATH}. The objective of these experiments (i.e., case studies) was to verify that the dynamic loader indeed rejects attacker-controlled shared objects before execution is transferred to application-level code.

\subsection{Case Study 1}

In the first experiment, an attacker-controlled shared object was placed in a writable directory (\texttt{/tmp}) using the same filename and exported symbols as a legitimate dependency. The malicious library preserved the expected ABI and implemented the required symbols so that the target application could continue execution without immediate failures or symbol resolution errors. The attacker then manipulated the dynamic loader's search precedence using \texttt{LD\_LIBRARY\_PATH}, causing the loader to resolve the malicious library before the legitimate system dependency. Without enforcement enabled, the dynamic loader accepted the attacker-controlled shared object as a valid dependency and mapped it into the application's address space during the runtime linking phase. As a result, attacker-controlled code executed transparently within the context of the victim process, while the application itself continued operating normally from the user's perspective. This behavior was confirmed experimentally by observing that the hijacked library executed successfully and produced the attacker-defined output (\texttt{*** HIJACKED ***}) during program execution.

We then enabled our proposed path-based enforcement mechanism using the \texttt{LD\_AUDIT} interface. During shared-object resolution, the verifier compared the resolved canonical path against the authenticated
whitelist manifest and rejected the unauthorized library before application execution proceeded. The verifier terminated execution because the attacker-controlled library path did not correspond to an approved whitelist
entry.

\subsection{Case Study 2}
The second experiment evaluated the Build-ID-based enforcement variant. As in the previous experiment, the attacker manipulated \texttt{LD\_LIBRARY\_PATH} to force resolution of a malicious shared object from \texttt{/tmp}. Without enforcement, the malicious library was loaded successfully. However, once Build-ID-based verification was enabled, the verifier extracted the ELF Build-ID and computed the SHA-256 digest of the resolved object. Since the attacker-controlled library's message digest was not present in the authenticated manifest, the verifier rejected the object and terminated execution before the library could be accepted as a valid runtime dependency.

\begin{figure}[t]
    \centering
    \includegraphics[width=\linewidth]{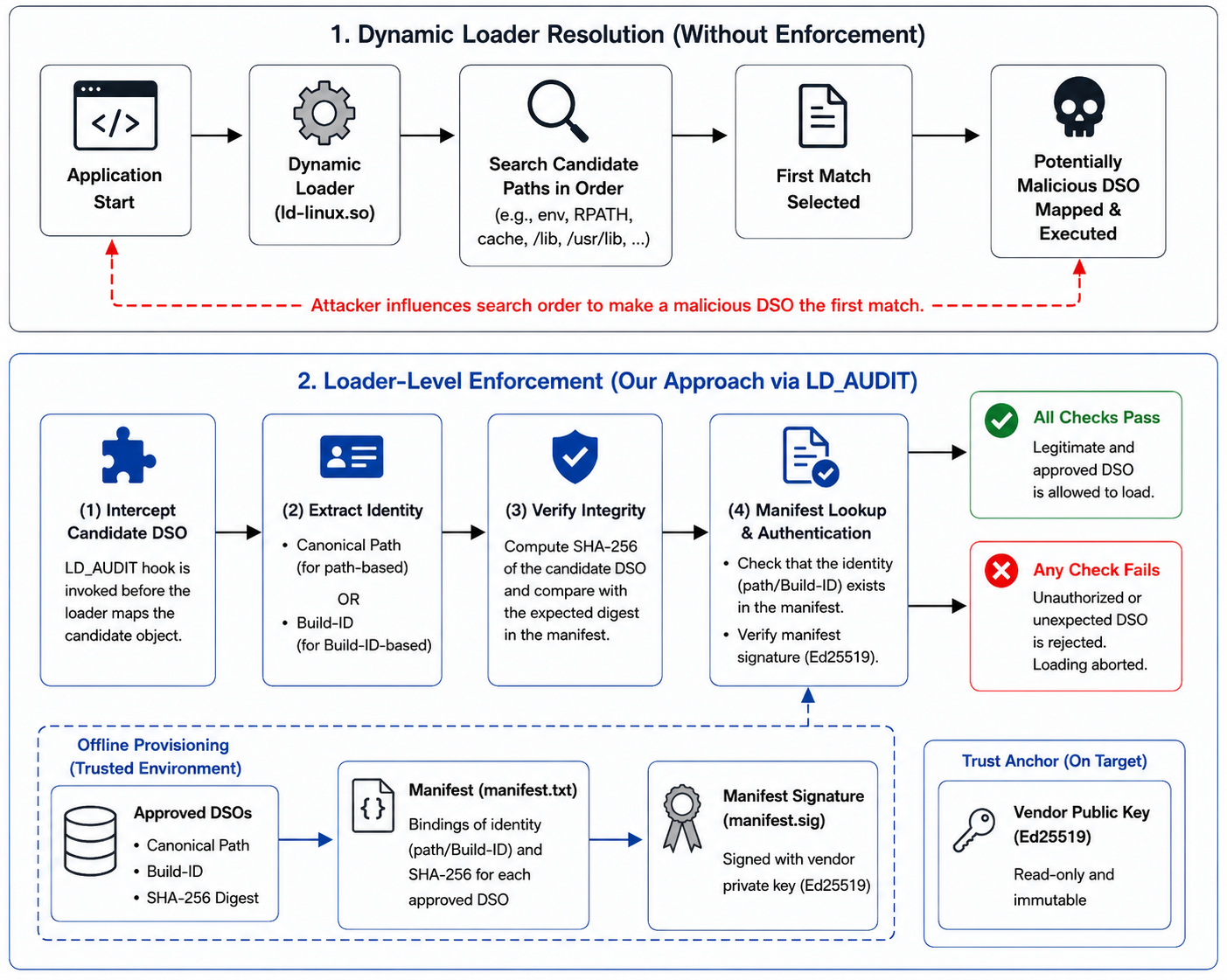}
    \caption{Steps for case studies}
    \label{fig:security-analysis}
\end{figure}

These experiments, summarized in Table~\ref{tab:hijack-validation}
and illustrated in Figure~\ref{fig:security-analysis},
demonstrate that the proposed mechanism successfully prevents
runtime shared-object hijacking attacks that exploit dynamic
loader search precedence\cite{anonymous2}.

\begin{table}[h]
\centering
\caption{Hijacking prevention validation results}
\label{tab:hijack-validation}

\begin{tabular}{lc}
\hline
Configuration & Result \\
\hline

No Enforcement & Hijack succeeds \\
Path-based Enforcement & Blocked \\
Build-ID-based Enforcement & Blocked \\

\hline
\end{tabular}
\end{table}

\section{Discussion}

Securing Linux systems against shared library hijacking remains a challenging problem due to the fundamental design of dynamic linking and the assumptions made by existing security mechanisms. Unlike traditional code-integrity attacks, shared-library hijacking does not require modifying trusted binaries or libraries. Instead, adversaries exploit the dynamic linker’s resolution logic by manipulating search paths, which enables malicious shared objects to be loaded in place of legitimate ones. This subtlety makes hijacking attacks difficult to detect and even harder to prevent using conventional integrity enforcement approaches.

\subsection{Practical Implications Across Platforms}

Our approach works for both general-purpose and embedded Linux environments. The glibc-based implementation shows that existing loader extension points can be leveraged to enforce verification without invasive system changes. This cross-platform applicability is significant, as embedded systems often lack hardware-backed integrity mechanisms and operate under constrained security assumptions.

These results suggest that loader-aware defenses provide a practical middle ground between heavyweight system-wide integrity frameworks and brittle binary-level hardening techniques. In environments where firmware images are static and supply chains are well-controlled, the proposed approach provides robust protection against runtime hijacking.

To further clarify the distinction between file integrity enforcement and correctness of resolution, we emphasize that kernel integrity solutions and our solution are orthogonal. Kernel-level integrity mechanisms such as IMA and fs-verity \cite{fsverity} provide strong guarantees regarding the authenticity and integrity of individual files. However, these systems verify file contents independently of dynamic linking semantics. Shared library hijacking is fundamentally a resolution correctness problem: the attack succeeds not because a file lacks integrity, but because the dynamic loader resolves and binds an unintended library that may nonetheless be validly signed. Consequently, even if all shared objects are protected by IMA or fs-verity, incorrect resolution order or search path manipulation can still result in loading a maliciously signed library. Our mechanism addresses this complementary problem by enforcing identity binding at the dynamic loader boundary. As the security properties differ, a direct performance comparison would not evaluate equivalent functionality.

\subsection{Why Build-ID Alone Is Insufficient}
\label{subsec:buildid-vs-sha256}

One could wonder why an additional cryptographic hash is required when ELF binaries already contain a Build-ID. 

The ELF Build-ID appears to provide a convenient integrity primitive, as it is commonly represented as a hexadecimal value resembling a SHA-1 hash. In many toolchains, the default Build-ID generation method is indeed derived from a hash over selected linker inputs or internal linker state \cite{gnuldbuildid}. However, our experimental analysis revealed a critical limitation of relying solely on the Build-ID for integrity enforcement. More specifically, we observed that patching an existing binary compiled with glibc, for example, by modifying executable sections or injecting code, does not necessarily result in a recomputed Build-ID. In practice, the Build-ID is typically generated once during the linking stage and embedded into the ELF object as static metadata. Subsequent post-build modifications performed directly on the binary file may therefore leave the original Build-ID unchanged. As a result, an attacker could alter the behavior of a shared object while preserving a previously trusted Build-ID value. This observation demonstrates that the Build-ID alone should not be interpreted as a sufficient cryptographic integrity guarantee. Instead, it is more accurately viewed as a stable artifact identifier that can distinguish builds and correlate binaries with associated metadata. Therefore, our approach combines Build-ID–based identification (as a primary key in the manifest) with SHA-256 hashing to ensure that both object identity and full byte-level integrity are verified during runtime loading. Hence, these observations demonstrate that the Build-ID alone is insufficient as an integrity mechanism. For this reason, our design treats the Build-ID solely as a primary key in our manifest that selects the intended artifact, and delegates integrity verification to a separate cryptographic hash.




 \subsection{Relationship to Software Supply Chain Security}

Modern software supply-chain security mechanisms primarily focus on protecting the integrity and provenance of software artifacts during development, packaging, and distribution \cite{newman2022sigstore, torres2019toto, slsa}. Examples include signed packages, reproducible builds, Software Bill of Materials (SBOM) frameworks, and integrity verification systems such as IMA and fs-verity. These mechanisms help ensure that software components originate from trusted sources and have not been tampered with during delivery or installation.

However, shared library hijacking is a distinct class of threat that occurs at runtime during dynamic linking. In such attacks, attackers do not necessarily modify trusted binaries or compromise the build pipeline itself. Instead, they manipulate the dynamic loader’s resolution process so that an unintended shared object is selected and loaded before the legitimate dependency. As a result, even correctly signed and authentic software packages may still be vulnerable if runtime library resolution can be influenced through search-path manipulation or unsafe loading semantics.

Our approach, therefore, complements existing software supply-chain protections~\cite{newman2022sigstore,torres2019toto,slsa}. Supply-chain integrity mechanisms establish trust in distributed artifacts, whereas our mechanism enforces runtime authenticity of dynamically resolved shared objects after deployment. More specifically, the proposed loader-level verification ensures that only explicitly approved libraries can be resolved and mapped into the process address space, even if an attacker can place alternative libraries within the loader’s search scope.

This distinction is necessary because traditional integrity systems verify whether a library is authentic, but generally do not verify whether it is the correct library intended for a particular runtime dependency resolution event. In fact, by binding runtime-loaded shared objects to approved identities and cryptographic hashes, our mechanism constrains a class of post-deployment dependency substitution attacks that traditional supply-chain protections may not address.

It should be noted that our mechanism does not defend against all software supply chain attacks at the same time, for example, compromise of the trusted build environment itself. If an attacker successfully compromises the software supply chain prior to deployment, for example, by injecting malicious code into the build pipeline or producing malicious libraries that are subsequently signed and approved, then the loader-level verification process would still accept those artifacts. Therefore, our work should be viewed as a complementary runtime enforcement mechanism that strengthens post-deployment guarantees of dependency authenticity within a broader software supply-chain security architecture.

\subsection{Limitations and Future Directions}

Our prototype also comes with certain limitations. The presented prototype relies on glibc’s \texttt{LD\_AUDIT} mechanism and therefore inherits its limitations. More specifically, audit modules may be disabled in secure-execution contexts such as setuid binaries, and the approach is specific to glibc-based systems. Furthermore, if the underlying embedded ecosystem is compiled with the root user, the initial access will be in the rights of the root user; then an attacker can modify the verifier or its trust anchor, and integrity guarantees no longer hold. Therefore, it is essential to separate the root user and the regular users.

The verification overhead of our mechanism scales approximately linearly with both the number and size of dynamically loaded shared objects. For each resolved library, the verifier performs filesystem access, ELF metadata parsing, identity extraction, and SHA-256 hashing prior to transferring control to the application. Our evaluation indicates that cryptographic hashing is the dominant cost component, while Build-ID extraction introduces negligible additional overhead. Therefore, applications with large dependency graphs, plugin-heavy architectures, or highly dynamic loading behavior may experience increased process initialization latency. This effect is particularly visible for short-lived command-line utilities, where startup overhead dominates total execution time. In contrast, for long-running services, embedded appliance-style deployments, and continuously executing user-space applications, the practical impact decreases over long execution durations. 


Our mechanism assumes that the manifest and its corresponding signatures are generated from a trusted build environment. If the build pipeline itself is compromised through software supply chain threats, an attacker could produce malicious libraries along with matching Build-IDs and cryptographic hashes, which would then be white-listed during provisioning. In such a scenario, the loader-level enforcement would validate the compromised artifacts, as the identity and integrity checks would still succeed. Therefore, the proposed system protects against unauthorized runtime substitution and relocation attacks but does not defend against supply-chain compromise at build time.

Future work would explore tighter integration with build systems to automate manifest generation, support for additional executable formats, and broader compatibility with containerized and sandboxed execution environments. Additionally, we aim to extend our approach to the musl loader.

\section{Conclusion}

Shared library hijacking exploits the dynamic linker’s resolution semantics rather than directly modifying trusted binaries, making it difficult to mitigate with existing security mechanisms. Integrity-based approaches, such as Linux IMA, are effective against file tampering but do not constrain which shared objects are selected at runtime. As a result, they fail to prevent attackers from injecting malicious libraries through search-order manipulation during shared object loading.

In this work, we showed that shared library hijacking is fundamentally a resolution correctness problem rather than a traditional integrity issue. To address this gap, we proposed a loader-level verification mechanism that enforces a whitelist of shared object identity during dynamic linking. 


We evaluated the proposed mechanism on both general-purpose and embedded Linux systems, demonstrating its practicality across different Linux deployment contexts. The results demonstrate that loader-aware verification can effectively prevent shared library hijacking attacks that can be deployed across desktops, servers, and firmware-centric embedded environments.


\section{Reproducibility}

To support reproducibility and independent verification of our results, we provide the complete implementation, experimental artifacts, evaluation scripts, and attack demonstrations through an anonymous GitHub repository. The repository also includes screenshots \cite{anonymous2}, proof-of-concept hijacking scenarios, and instructions for reproducing the experimental setup and the evaluation results presented in this paper.

The anonymous artifact repository is available at \cite{anonymous}.

\section{Acknowledgement}
This work was supported by the Flemish Government through the Cybersecurity Research Program.

\bibliographystyle{plain}
\bibliography{references}

@inproceedings{1_payer2012safe,
  title={Safe loading-a foundation for secure execution of untrusted programs},
  author={Payer, Mathias and Hartmann, Tobias and Gross, Thomas R},
  booktitle={2012 IEEE Symposium on Security and Privacy},
  pages={18--32},
  year={2012},
  organization={IEEE}
}

@inproceedings{2_ge2017evil,
  title={An Evil Copy: How the Loader Betrays You.},
  author={Ge, Xinyang and Payer, Mathias and Jaeger, Trent},
  booktitle={NDSS},
  year={2017}
}

@book{4_williams2021improving,
  title={Improving Security Through Egalitarian Binary Recompilation},
  author={Williams-King, David},
  year={2021},
  publisher={Columbia University}
}

@article{5_wheeler2001secure,
  title={Secure Programming for Linux and Unix HOWTO},
  author={Wheeler, David A},
  journal={http://www. linux. org/docs/ldp/howto/Secure-Programs-HOWTO/index. html},
  year={2001}
}

@inproceedings{6_di2015elf,
  title={How the $\{$ELF$\}$ Ruined Christmas},
  author={Di Federico, Alessandro and Cama, Amat and Shoshitaishvili, Yan and Kruegel, Christopher and Vigna, Giovanni},
  booktitle={24th USENIX Security Symposium (USENIX Security 15)},
  pages={643--658},
  year={2015}
}

@inproceedings{7_park2005new,
  title={A NEW MECHANISM FOR OS SECURITY: Selective Checking of Shared Library Calls for Security},
  author={Park, Dae Yeon},
  booktitle={WEB Information Systems and Technologies},
  pages={381--388},
  year={2005}
}

@article{8_carbone2014malware,
  title={Malware Memory Analysis of the Jynx2 Linux Rootkit (Part 1): Investigating a Publicly Available Linux Rootkit Using the Volatility Memory Analysis Framework},
  author={Carbone, Richard},
  year={2014}
}

@book{9_vijayakumar2014protecting,
  title={Protecting programs during resource access},
  author={Vijayakumar, Hayawardh},
  year={2014},
  publisher={The Pennsylvania State University}
}

@phdthesis{10_zakaria2025exploiting,
  title={Exploiting Stability in Software Systems: Primitives for Fast Startup, Binary Introspection, and Explicit Dependency Control},
  author={Zakaria, Farid},
  year={2025},
  school={University of California, Santa Cruz}
}

@misc{13_ditullio2020context,
  title={Context check bypass to enable opening shared-object libraries},
  author={DiTullio, Jeff and Fenton, Michael Ryan and Koppel, James Brandon and Lundeen, Timothy D},
  year={2020},
  month=mar # "~31",
  publisher={Google Patents},
  note={US Patent 10,606,612}
}

@article{14_kwon2011automatic,
  title={Automatic detection of unsafe dynamic component loadings},
  author={Kwon, Taeho and Su, Zhendong},
  journal={IEEE Transactions on Software Engineering},
  volume={38},
  number={2},
  pages={293--313},
  year={2011},
  publisher={IEEE}
}

@inproceedings{15_castes2023dynamic,
  title={Dynamic Linkers Are the Narrow Waist of Operating Systems},
  author={Castes, Charly and Ghosn, Adrien},
  booktitle={Proceedings of the 12th Workshop on Programming Languages and Operating Systems},
  pages={26--33},
  year={2023}
}

@inproceedings{16_larose2023dynamic,
  title={Dynamic Library Compartmentalization},
  author={Larose, Octave},
  booktitle={Companion Proceedings of the 2023 ACM SIGPLAN International Conference on Systems, Programming, Languages, and Applications: Software for Humanity},
  pages={51--52},
  year={2023}
}

@article{beazley2001inside,
  title={The inside story on shared libraries and dynamic loading},
  author={Beazley, David M and Ward, Brian D and Cooke, Ian R},
  journal={Computing in Science \& Engineering},
  volume={3},
  number={5},
  pages={90--97},
  year={2001},
  publisher={IEEE}
}

@misc{17_IMA,
  author = {{Linux Integrity Subsystem}},
  title = {Integrity Measurement Architecture (IMA)},
  howpublished = {\url{https://ima-doc.readthedocs.io/en/latest/ima-concepts.html}},
  year = {2026}
}

@misc{anonymous,
  author = {anonymous},
  title = {{anonymous}},
  howpublished = "\url{https://anonymous.4open.science/r/shareb\_object\_library\_hijacking-5E36/}", 
  note = "[Online; accessed 25-05-2026]"
}

@misc{anonymous2,
  author = {anonymous2},
  title = {{anonymous2}},
  howpublished = "\url{https://anonymous.4open.science/r/shareb_object_library_hijacking-5E36/PathBasedEnforcement/CaseStudies_AttackPrevention/README.md}",
  note = "[Online; accessed 25-05-2026]"
}

@misc{resolutionorder,
  author = {Michael Kerrisk},
  title = {ld.so(8) Linux Programmer's Manual},
  howpublished = {\url{https://man7.org/linux/man-pages/man8/ld.so.8.html}},
  year = {2026}
}

@misc{musl,
  author = {Rich Felker},
  title = {musl libc},
  howpublished = {\url{https://musl.libc.org}},
  year = {2026}
}

@misc{hyperfine,
  author = {sharkdp},
  title = {hyperfine: A command-line benchmarking tool},
  howpublished = {\url{https://github.com/sharkdp/hyperfine}},
  year = {2026}
}

@misc{fsverity,
  title = {fs-verity: read-only file-based authenticity protection},
  author = {{Linux Kernel Documentation}},
  year = {2024},
  howpublished = {\url{https://docs.kernel.org/filesystems/fsverity.html}}
}

@inproceedings{torres2019toto,
  title={in-toto: Providing farm-to-table guarantees for bits and bytes},
  author={Torres-Arias, Santiago and Afzali, Hammad and Kuppusamy, Trishank Karthik and Curtmola, Reza and Cappos, Justin},
  booktitle={28th USENIX Security Symposium (USENIX Security 19)},
  pages={1393--1410},
  year={2019}
}

@misc{slsa,
  title = {SLSA: Supply-chain Levels for Software Artifacts},
  author = {{OpenSSF}},
  year = {2021},
  howpublished = {\url{https://slsa.dev}}
}

@inproceedings{agrawal2015architectural,
  title={Architectural support for dynamic linking},
  author={Agrawal, Varun and Dabral, Abhiroop and Palit, Tapti and Shen, Yongming and Ferdman, Michael},
  booktitle={Proceedings of the Twentieth International Conference on Architectural Support for Programming Languages and Operating Systems},
  pages={691--702},
  year={2015}
}

@inproceedings{newman2022sigstore,
  title={Sigstore: Software signing for everybody},
  author={Newman, Zachary and Meyers, John Speed and Torres-Arias, Santiago},
  booktitle={Proceedings of the 2022 ACM SIGSAC Conference on Computer and Communications Security},
  pages={2353--2367},
  year={2022}
}

@misc{gnuldbuildid,
  author = {{GNU Project}},
  title = {GNU Linker Options: --build-id},
  howpublished = {\url{https://sourceware.org/binutils/docs/ld/Options.html}},
  year = {2026}
}

@misc{glibc,
  author = {{GNU Project}},
  title = {GNU C Library (glibc)},
  year = {2026},
  howpublished = {\url{https://www.gnu.org/software/libc/}}
}

@misc{rtldaudit,
  author = {Michael Kerrisk},
  title = {rtld-audit(7) Linux Programmer's Manual},
  howpublished = {\url{https://man7.org/linux/man-pages/man7/rtld-audit.7.html}},
  year = {2026}
}

@misc{elfspec,
  title = {System V Application Binary Interface: ELF gABI},
  howpublished = {\url{https://refspecs.linuxfoundation.org/elf/gabi4+/contents.html}}
}

@phdthesis{18_lu2017securing,
  title={Securing software systems by preventing information leaks.},
  author={Lu, Kangjie},
  year={2017},
  school={Georgia Institute of Technology, Atlanta, GA, USA}
}

@article{19_silakov2012using,
  title={Using virtualization to protect application address space inside untrusted environment},
  author={Silakov, Denis V},
  journal={Programming and Computer Software},
  volume={38},
  number={1},
  pages={24--33},
  year={2012},
  publisher={Springer}
}

@inproceedings{22_zhang2013secgot,
  title={Secgot: Secure global offset tables in elf executables},
  author={Zhang, Chao and Duan, Lei and Wei, Tao and Zou, Wei},
  booktitle={Conference of the 2nd International Conference on Computer Science and Electronics Engineering (ICCSEE 2013)},
  pages={995--998},
  year={2013},
  organization={Atlantis Press}
}

@article{23_goonasekera2015libvm,
  title={LibVM: an architecture for shared library sandboxing},
  author={Goonasekera, Nuwan and Caelli, William and Fidge, Colin},
  journal={Software: Practice and Experience},
  volume={45},
  number={12},
  pages={1597--1617},
  year={2015},
  publisher={Wiley Online Library}
}

@article{24_jeong2020cfi,
  title={A cfi countermeasure against got overwrite attacks},
  author={Jeong, Seunghoon and Hwang, Jaejoon and Kwon, Hyukjin and Shin, Dongkyoo},
  journal={IEEE Access},
  volume={8},
  pages={36267--36280},
  year={2020},
  publisher={IEEE}
}

@article{25_jianjun2022defense,
  title={Defense method against code reuse attack based on real-time code loading and unloading},
  author={Jianjun, HUANG and LIANG, Bin and others},
  journal={Computer Science},
  volume={49},
  number={10},
  pages={279--284},
  year={2022}
}

@inproceedings{26_zhang2015control,
  title={Control flow and code integrity for COTS binaries: An effective defense against real-world ROP attacks},
  author={Zhang, Mingwei and Sekar, R},
  booktitle={Proceedings of the 31st Annual Computer Security Applications Conference},
  pages={91--100},
  year={2015}
}

@article{27_fu2012dynamic,
  title={Dynamic detection of component loading vulnerability},
  author={Fu, Jianming and Peng, Bichen and Du, Hao},
  journal={Journal of Tsinghua University Science and Technology},
  volume={52},
  number={10},
  year={2012},
  publisher={Tsinghua University Press, Tsinghua University Beijing 100084 China}
}

@inproceedings{28_wang2018lprov,
  title={Lprov: Practical library-aware provenance tracing},
  author={Wang, Fei and Kwon, Yonghwi and Ma, Shiqing and Zhang, Xiangyu and Xu, Dongyan},
  booktitle={Proceedings of the 34th Annual Computer Security Applications Conference},
  pages={605--617},
  year={2018}
}

@inproceedings{29_rommel2023thread,
  title={Thread-level attack-surface reduction},
  author={Rommel, Florian and Dietrich, Christian and Ziegler, Andreas and Ostapyshyn, Illia and Lohmann, Daniel},
  booktitle={Proceedings of the 24th ACM SIGPLAN/SIGBED International Conference on Languages, Compilers, and Tools for Embedded Systems},
  pages={64--75},
  year={2023}
}

@inproceedings{30_porter2020blankit,
  title={Blankit library debloating: Getting what you want instead of cutting what you don’t},
  author={Porter, Chris and Mururu, Girish and Barua, Prithayan and Pande, Santosh},
  booktitle={Proceedings of the 41st ACM SIGPLAN Conference on Programming Language Design and Implementation},
  pages={164--180},
  year={2020}
}

\end{document}